\documentclass[ras]{agutex}

\usepackage{graphicx}
\usepackage{amssymb}

\authorrunninghead{HELMBOLDT ET. AL}

\titlerunninghead{VLSS-BASED TEC DISTURBANCE CLIMATOLOGY}

\authoraddr{J.\ F.\ Helmboldt, US Naval Research Laboratory, 
Code 7213, 4555 Overlook Ave. SW, Washington, DC 20375 
(wendy.peters@nrl.navy.mil)}

\authoraddr{W.\ M.\ Lane Peters, US Naval Research Laboratory, 
Code 7213, 4555 Overlook Ave. SW, Washington, DC 20375 
(joe.helmboldt@nrl.navy.mil)}

\authoraddr{W.\ D.\ Cotton, National Radio Astronomy Observatory, 
520 Edgemont Rd, Charlottesville, VA 22903 
(bcotton@nrao.edu)}

\begin{document}

\title{Climatology of Mid-latitude Ionospheric Disturbances from the Very Large Array Low-frequency Sky Survey}

\authors{J. F. Helmboldt, \altaffilmark{1} W.\ M.\ Lane, \altaffilmark{1} \& W.\ D.\ Cotton \altaffilmark{2}}

\altaffiltext{1}{US Naval Research Laboratory, Washington, DC, USA.}
\altaffiltext{2}{National Radio Astronomy Observatory, Charlottesville, VA, USA.}

\begin{abstract}
The results of a climatological study of ionospheric disturbances derived from observations of cosmic sources from the Very Large Array (VLA) Low-frequency Sky Survey (VLSS) are presented.  We have used the ionospheric corrections applied to the 74 MHz interferometric data within the VLSS imaging process to obtain fluctuation spectra for the total electron content (TEC) gradient on spatial scales from a few to hundreds of kilometers and temporal scales from less than one minute to nearly an hour.  The observations sample nearly all times of day and all seasons.  They also span latitudes and longitudes from $28^\circ$N to $40^\circ$N and $95^\circ$W to $114^\circ$W, respectively.  We have binned and averaged the fluctuation spectra according to time of day, season, and geomagnetic ($K_p$ index) and solar ($F10.7$) activity.  These spectra provide a detailed, multi-scale account of seasonal and intraday variations in ionospheric activity with wavelike structures detected at wavelengths between about 35 and 250 km.  In some cases, trends between spectral power and $K_p$ index and/or $F10.7$ are also apparent.  In addition, the VLSS observations allow for measurements of the turbulent power spectrum down to periods of 40 seconds (scales of $\sim\!0.4$ km at the height of the $E$-region).  While the level of turbulent activity does not appear to have a strong dependence on either $K_p$ index or $F10.7$, it does appear to be more pronounced during the winter daytime, summer nighttime, and near dusk during the spring.
\end{abstract}

\begin{article}

\section{Introduction}
The ionosphere is known to contain a host of transient ionospheric disturbances.  They range from fine-scale, turbulent like density fluctuations \citep[e.g.,][]{coh09} to large-scale plasma waves \citep[e.g.,][]{bor09} and large areas of storm-enhanced density \citep[e.g.,][]{fos02}.  Some of the most common among these are so called traveling ionospheric disturbances (TIDs) which are known to be plasma density waves with a range of sizes and speeds.  TIDs are often divided into two classes, large-scale TIDs (LSTIDs) and medium-scale TIDs (MSTIDs).  LSTIDs are typically associated with geomagnetic storms and have sizes $\sim\!2000$ km, periods of $\sim\!\!1$ hour, and speeds of about 700 m s$^{-1}$ \citep{bor09}.  MSTIDs have periods between 10 minutes and $\sim\!1$ hour, speeds between 50 and a few hundred m s$^{-1}$, and sizes $\sim\!100$ km \citep[e.g.,][]{jac92,jac95}.\par
Unlike their large-scale counterparts, MSTIDs are not typically associated with geomagnetic storms and are likely generated by more localized processes.  While the climatology of these relatively commonly occurring waves has been studied in the past \citep[see, e.g.,][]{her06,tsu07}, the mechanisms responsible for generating them are still not clear.  They are most frequently seen at dawn and dusk, implying a relationship to the dynamics associated with the transition between the daytime and nighttime ionospheres.  In the northern hemisphere, daytime MSTIDs are seen most often in the winter, usually propagating toward the southeast \citep{her06}.  MSTIDs that appear during the night are most prevalent in the summer, mostly directed toward the southwest \citep{her06,tsu07}.  The nighttime MSTIDs in particular remain something of a mystery, especially regarding their orientation and direction of motion.  Their formation may be related to the Perkins instability \citep{per73}, and/or to coupling between instabilities within the $E$ and $F$ regions \citep[see, e.g.,][]{cos04}.  Different mechanisms have been proposed to explain their southwestward motion \citep{kel01,cos07,yok09}, however, none have been conclusively verified by observations.\par
In addition, MSTIDs have been observed to propagate in a variety of other directions, attributed to changes in the $F$-region neutral wind and/or orographic gravity waves \citep[e.g.,][]{shi08,dym11,hel12b} and to deposition of material from the plasmasphere \citep{helint}.  Both of these types of MSTIDs in particular can have relatively weak amplitudes, between $10^{-3}$ and $10^{-2}$ TECU (1 TECU$=\!10^{16}$ m$^{-2}$).  The presence (or lack thereof) of MSTIDs may also be linked to increased levels of turbulent activity within the ionosphere \citep[e.g.,][]{hel12b}.\par
A detailed understanding of the generation, maintenance, and evolution of MSTIDs requires a climatological study which is sensitive to the full range of observable MSTIDs and to smaller-scale density fluctuations to which they may be related.  Recently, \citet{hel12a,hel12b} have demonstrated that radio interferometers, especially those observing in the VHF regime, may be powerful tools for producing such a study.  With the capability to measure differential total electron content (TEC) with a precision $<\! 10^{-3}$ TECU on spacings of 1 km or less when observing a bright cosmic source \citep{coh09,hel12a}, such interferometers are capable of observing small TEC fluctuations such as those related to turbulence.  Using long temporal baselines, \citet{hel12a} showed that one can also detect and characterize the properties of MSTIDs as they pass over the interferometer, some of which have relatively small fluctuation amplitudes ($\sim\! 10^{-3}$ TECU).  In addition, \citet{helint} have demonstrated that by using simultaneous observations of several moderately bright VHF sources, one can characterize fluctuations on much larger scales.  For instance, at 74 MHz, the $15^\circ$ field of view of the Karl G.\ Jansky Very Large Array (VLA) in New Mexco corresponds to an area 80 km in diameter at a height of 300 km.\par
Here, we seek to apply these radio interferometer-based techniques to a large, 74 MHz survey conducted with the VLA to obtain a statistical, climatological picture of ionospheric disturbances spanning a wide range of scales in size, oscillation period, and amplitude.  In terms of amplitude and size, this study constitutes one of the largest dynamic range investigations of its kind, and it contains one of the first climatological studies of ionospheric turbulence.  In Section 2, we briefly describe the characteristics of the survey used, the VLA Low-frequency Sky Survey \citep[VLSS;][]{coh07}.  In the next section, we detail the analysis used and ionospheric data products that were generated.  In Section 4, we show the results of our climatological study and discuss them further in Section 5.

\section{Experimental Detail}
In order to create high quality astronomical images, the VLSS standard data processing corrects the data for ionospheric effects.  In this study, we re-analyze the data products from that processing to extract information about the dynamics of the ionosphere.
\subsection{The Survey}
The VLSS is a survey of the northern sky covering declinations north of about $-30^\circ$.  It was conducted between 2001 and 2007 using the 74 MHz system of the VLA \citep{kas07} with full dual polarization (VLA program numbers AP397, AP441, AP452, and AP509).  The VLA was used in its B-configuration, an inverted ``Y'' of 27 antennas spanning roughly 11 km.  Observations of sources with low declinations were conducted with the VLA in the hybrid ``BnA'' configuration which has the northern arm extended to its A-configuration size of nearly 20 km.  Hybrid configurations such as BnA are routinely used when observing sources at low elevations to improve the uniformity of the antenna positions projected onto the sky.  Approximately 20\% of the VLSS observations were conducted in the BnA configuration.\par
Other VLA configurations, which are available roughly four months at a time, are the A, C, and D configurations spanning roughly 40, 4, and 1 km, respectively.  Observations made at 74 MHz with the two most compact configurations have a substantially larger effective system temperature due to the fact that their shorter baselines make them more sensitive to diffuse emission from the Galaxy, generally referred to as ``sky noise.''  The Galactic emission is dominated by synchrotron emission that is much brighter at lower frequencies and on relatively large angular scales.  Consequently, the A and B configurations have lower sky noise contributions due to the fact that their smallest baselines are too big to be sensitive to the large-angular-scale Galactic emission.  The B configuration was chosen for the VLSS because its angular resolution, 80 arcseconds, is similar to that of other northern sky surveys at 325 MHz \citep[54 arcseconds;][]{ren97} and 1400 MHz \citep[45 arcseconds;][]{con98} to facilitate the compilation of radio-frequency spectra for detected sources.\par
The details of the survey and associated data processing are discussed in detail by \citet{coh07}.  Recently, improvements in imaging and calibration techniques as well as new algorithms for mitigating radio frequency interference (RFI) were combined to re-process all of the VLSS data.  The results of this effort are referred to as ``VLSS redux,'' or VLSSr, and the improvements over the previously used processing techniques are discussed by \citet{lan12}.\par
Briefly, the VLSS mapped the sky using 523 overlapping pointings and imaging the entire $15^\circ$ wide, circular field of view of the VLA at 74 MHz at each pointing.  Each pointing was observed for a total of about 1.25 hours or more, broken into four or more 20-minute segments, or three or more 25-minute segments.  For each pointing, these segments were spaced by hours or days.  This amounted to a total of 1862 separate $\sim\! 20$-minute observations.  These observations cover nearly all times of day and season and the associated ionospheric pierce-points span a range in latitude and longitude, which we have illustrated in Fig.\ \ref{cover}.
\subsection{Field-based Calibration}
The effect of the ionosphere on the VLSS observations varied significantly across the $15^\circ$ field of view, which corresponds to a width of 80 km at an altitude of 300 km.  The first-order effect of the ionosphere is to shift the positions of observed sources by an amount proportional to the mean TEC gradient over the area spanned by the array along the line of sight to each source.  The VLSS used a procedure referred to as ``field-based'' calibration \citep{cot04} to account for these shifts as a function of time and position within the field of view.  This calibration scheme uses ``snapshot'' images of relatively bright sources using 1--2 minutes of data for each snapshot.  The positions of these sources are then measured within the snapshots and compared to a reference catalog.  The original VLSS used a VLA-based, 1400 MHz catalog \citep{con98} as its reference.  This produced a catalog of 74 MHz sources whose positions are tied to the bright sources contained within the 1400 MHz reference catalog.\par
Substantial improvements were obtained in ionospheric calibration by \citet{lan12} for the VLSSr by using the original VLSS as the reference catalog.  In other words, \citet{lan12} essentially produced a second iteration of the field-based imaging scheme where the catalog of 74 MHz sources generated by the first iteration was used as the reference catalog during the second iteration.  Since there is a variety of spectral shapes among cosmic radio sources, not all bright 1400 MHz sources are also bright at 74 MHz, and vice versa.  Consequently, having a reference catalog of sources known to be bright (here, $>\!2.5$ Jy; 1 Jy $=10^{-26}$ W m$^{-2}$ Hz$^{-1}$) at 74 MHz made the field-based calibration process more robust in that a larger area could be searched within each snapshot for the shifted source position.
\subsection{Self-calibration}
While not used as part of the VLSS, a separate calibration technique referred to as ``self-calibration'' \citep{cor99} can be used to derive ionospheric corrections toward a single bright cosmic source if the intensity of that source dominates the field of view.  Self-calibration uses a model for the brightness distribution on the sky (often a simple point-source) to solve for antenna-based phase corrections.  For $N$ antennas, there are $N(N-1)/2$ redundant pairs or baselines, making this an over-determined problem.  The derived phase corrections are proportional to the difference in TEC, or ``$\delta \mbox{TEC}$'', along the lines of sight of the antennas and an arbitrarily chosen reference antenna.  \citet{hel12a} showed that using a bright source with the VLA at 74 MHz, these $\delta \mbox{TEC}$ measurements can be made to a precision of about $3\times 10^{-4}$ TECU, allowing for an analysis of very fine-scale, small amplitude fluctuations along the lines of sight toward a particular source.\par
To make such fine-scale ionospheric measurements, we sought to apply self-calibration whenever possible to the VLSS observations to complement the wide-field, medium-scale measurements provided by the field-based calibration process.  To do this, we had to first determine how bright a source needed to be to accurately compute $\delta \mbox{TEC}$ using self-calibration-derived phase corrections.  We did this by first identifying all sources in the VLSS catalog with peak intensities $>\! 6$ Jy beam$^{-1}$.  We then used each source to perform phase-only self-calibration (i.e., we did not solve for the amplitudes of the complex antenna gains), assuming a simple point-source model.  Since we would like to probe the smallest scales in space and time, we did this using the smallest time interval possible.  For the VLSS data, this was 10--20 seconds.  We then converted these self-calibration phases to $\delta \mbox{TEC}$ measurements by unwrapping the phases within each $\sim\!\! 20$ minute observations and de-trending with a linear fit.  This was done separately for each polarization so that we could use the rms difference between the de-trended $\delta \mbox{TEC}$ measurements from the two polarizations to assess the precision of the measurements.\par
\begin{figure}
\noindent\includegraphics[width=6in]{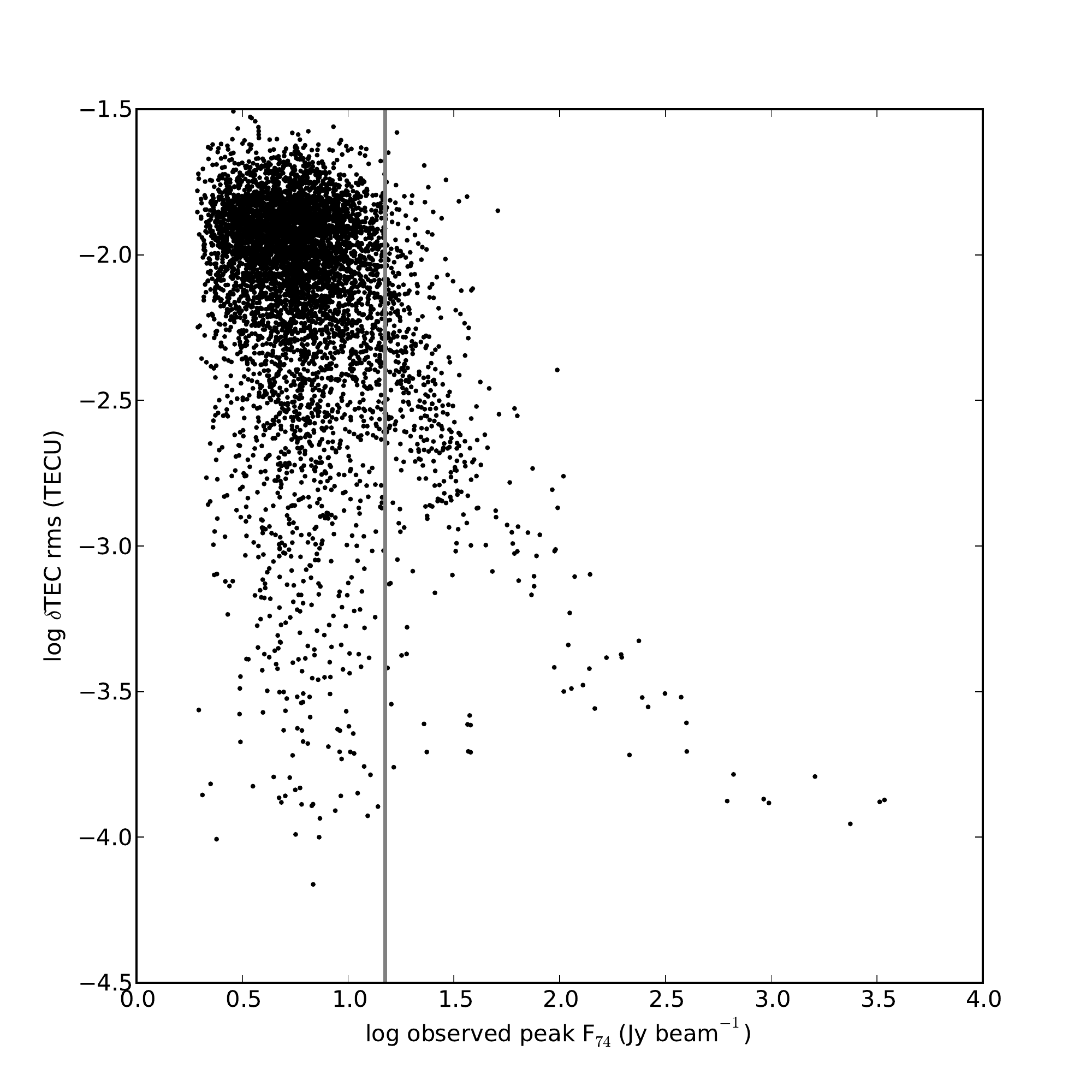}
\caption{The rms difference in $\delta \mbox{TEC}$ between the two polarizations among all baselines and time steps for each observation of each VLSS source brighter than 6 Jy versus its observed peak intensity (i.e., modified by the antenna response).  The vertical grey line shows the chosen limit of an intensity of $>\!15$ Jy beam$^{-1}$ for single-source data.  See Section 2.2 for more discussion.}
\label{rms}
\end{figure}

In Fig.\ \ref{rms}, we have plotted $\delta \mbox{TEC}$ rms versus the observed peak intensity (i.e., modified by the antenna response) for these sources.  Above a limit of about 15 Jy beam$^{-1}$, there is a noticeable anti-correlation between this rms and the source intensity which is what one would expect if the $\delta \mbox{TEC}$ measurements represent real ionospheric fluctuations and not noise.  Below this limit, the data essentially form a scatter plot, and the self-calibration phases are likely dominated by noise caused by ``confusion,'' i.e., the contribution from other sources in the field of view that were ignored in the self-calibration process.  Therefore, for the analysis that will be described below, we have only used sources with observed intensities $>\! 15$ Jy beam$^{-1}$.  There are 281 sources which meet this criterion that were observed within 477 different VLSS pointings (recall that the pointings overlapped).  These yielded 1978 separate observations of the 281 bright sources.\par
In addition to the 1978, 20-minute observations of relatively bright sources, the VLSS data also contain several five-minute observations of the extremely bright source Cygnus A, or ``CygA.''  At 74 MHz, Cyg A is brighter than 17000 Jy, making it an excellent source to use to determine corrections for instrumental effects which are relatively stable in time.  Consequently, Cyg A was observed 289 times during the VLSS campaign, each time with a duration of five minutes.  The extreme brightness of Cyg A also provides an excellent means for probing the smallest-scale/lowest-amplitude ionospheric fluctuations observable with the VLA \citep[see][]{hel12a,hel12b}.  Therefore, to increase the dynamic range of our study of ionospheric disturbances, we also applied self-calibration to all of the five-minute observations of Cyg A.  We note that because the Cyg A observations were shorter, the de-trending of the $\delta \mbox{TEC}$ measurements made from these observations biased them toward smaller-period oscillations.  This makes the Cyg A data complementary to the data produced from the 281 bright source observations described above, which are less sensitive to smaller-scale/smaller amplitude fluctuations because the sources used are not nearly as bright as Cyg A.

\section{Spectral Analysis}
\subsection{Multi-source Data}
The repository of source position shifts compiled during the processing of the VLSS provides a rich database of TEC gradient time series that can be analyzed and searched for evidence of wavelike activity.  Indeed, \citet{helint} showed that by Fourier transforming such a time series in time, and then in space, one may produce a three-dimensional (one temporal, two spatial) power spectrum ``cube'' of TEC gradient fluctuations.  In practice, a spectral cube is generated for both components (north-south and east-west) of the TEC gradient.  These two cubes are then added together such that the observed power of a wave with a TEC amplitude $A$ and spatial frequency $\xi$ will simply be $(2\pi \xi A)^2$.  These spectra are also normalized by the amplitude of the impulse response function (IRF), computed using the source positions projected to an ionospheric height of 300 km.  This is denoted in the units reported for these spectra by including a unit of IRF$^{-1}$.\par
The three-dimensional spectral cubes can be used to map the level and direction of any detected wavelike structures within the data.  We have applied the spectral analysis algorithm described in \citet{helint} to each roughly twenty-minute observation of each VLSS field using outputs from the VLSSr, yielding 1862 power-spectrum cubes.  Since each spectrum is based on position shift data from multiple sources, for the remainder of this paper, we will referred to these spectral data as ``multi-source'' data for convenience.\par
It was demonstrated by \citet{helint} that these spectra have excellent sensitivity, being able to detect fluctuations with amplitudes $<\! 10^{-3}$ TECU.  This is in part due to the improved data processing/calibration techniques detailed in \citet{lan12}.  They demonstrated the effect of these improvements on ionospheric analysis using mean fluctuation spectra from the entire survey.  They showed that the use of the original VLSS as the reference catalog as well as the new RFI mitigation techniques substantially improved the sensitivity of the power spectra to both large and small amplitude fluctuations.  Using a better reference catalog allows one to reliably search a larger area for each calibration source, facilitating the detection of large amplitude fluctuations.  Reducing the noise in each snapshot image by implementing new RFI-subtracting and flagging software allowed for the detection of smaller position shifts.  Specifically, the spatial frequency at which the mean fluctuation spectrum reached the noise ``floor'' was increased by a factor of two by the RFI-mitigation software, decreasing the typical wavelength of the weakest detectable disturbances from 70 to 35 km.

\subsection{Single-source Data}
While the power spectrum cubes described above provide a wealth of information about TEC fluctuations on a range of scales, their sensitivity is generally limited at relatively high spatial and temporal frequencies due to the effective smoothing applied to the data.  The position shifts of the calibrator sources give the mean TEC gradient over the span of the VLA, which in the case of the VLSS amounts to smoothing the data with an 11-km wide circular filter.  In addition, snapshots were made using 1--2 minutes of data, resulting in temporal smoothing.\par
However, the measurements of $\delta \mbox{TEC}$ made using individual bright sources described in Section 2.3 used temporal sampling of 10--20 seconds with mean antenna spacings $<\! 1$ km.  This implies that there is potentially more information within these data about fluctuations on scales too small to be detected with the multi-source spectra.  To help explore phenomena with smaller amplitudes and sizes, we sought to apply the methods developed by \citet{hel12a,hel12b} and refined by \citet{helint} to use 74 MHz VLA observations of a single bright source to characterize TEC fluctuations on significantly smaller scales.\par
The single-source spectral analysis is similar to that of its multi-source counterpart in that a three-dimensional Fourier transform of the TEC gradient is used.  However, in this case, the gradient is measured at each antenna for a single source rather than toward multiple sources.  Because of the Y-shape of the VLA, the $\delta \mbox{TEC}$ measurements obtained from self-calibration (see Section 2.3) cannot be used to directly compute the full two-dimensional TEC gradient at each antenna.  However, \citet{hel12a} showed that most of the structure in the TEC surface observed by the VLA can be recovered with a second-order polynomial fit to $\delta \mbox{TEC}$ as a function of antenna position.\par
Using these polynomial fits to estimate the gradient at each antenna, the same Fourier-based procedure can be applied as was used with the multi-source approach.  The details of how this is implemented are discussed at length by \citet{helint}.  There are two main difference between the spectral cubes produced from the single source data and those generated using multi-source data.  First, the spectral resolution in the spatial regime of the single-source-based spectra is much lower given the compact size of the VLA (11-km).  Because of this, each temporal mode within each spectral cube is well approximated by a single spatial frequency and direction.  Consequently, for each single-source spectrum, we computed a single peak power and weighted (with the spectral power) mean spatial frequency in the north-south, $\xi_{NS}$, and east-west, $\xi_{EW}$, directions for each temporal mode.  When combining all temporal frequencies together, this allows for a measurement of the distribution of fluctuation strength in the $\xi_{NS}$,$\xi_{EW}$ plane with much better spectral resolution than could normally be achieved with such a compact array.  This is because, by using a single spatial mode for each temporal mode, we have essentially converted the relatively long temporal baselines to spatial ones.  For instance, for a TID with a speed of 100 m s$^{-1}$, a 20-minute baseline amounts to a 120-km baseline, more than ten times larger than the B-configuration VLA.\par
The second difference between the single- and multi-source analyses is that because the cosmic sources are essentially infinitely far away, the lines of sight of the antennas toward a single source are basically parallel.  The lines of sight toward multiple sources, however, are far from parallel.  Thus, the physical separation among pierce-points for the multi-source observations is proportional to the assumed height, whereas for the single-source data, the pierce-point separations remain virtually the same for any assumed altitude.  This renders the multi-source data incapable of sensing disturbances in the upper ionosphere and plasmasphere that have wavelengths less than about 1000 km.  Conversely, the single-source observations are sensitive to $\sim\! 10$-km and larger sized fluctuations, evan at plasmaspheric heights \citep[see, e.g.,][]{helint}.\par
The single-source spectral analysis was applied to all 20-minute observations of the 281 identified bright sources and all five-minute observations of Cyg A (see Section 2.3).  In doing so, we found that for the observations conducted in the hybrid BnA configuration (see Section 2.1), the extended, 20-km long northern arm of the VLA posed a problem for the polynomial fit-based analysis.  In short, the polynomial fits only provide a uniform estimate of the TEC gradient across the array when it is symmetric.  Fitting a two-dimensional polynomial to an asymmetric array can lead to biasses within the spectral analysis performed using the results from such polynomial fits.  For instance, we found that data from this configuration produced two-dimensional fluctuation spectra that were, on average, elongated along the north/south axis.  Because of this inherent bias, single-source BnA observations (360 for the bright sources; 87 for Cyg A), were excluded from this analysis.\par
In addition to this three-dimensional spectral analysis, we also used the single source data to yield a statistical description of any isotropic fluctuations on the finest scales observable within the VLSS data.  This was done according to the prescription of \citet{hel12a} for using the self-calibration-determined $\delta \mbox{TEC}$ values to numerically compute the projection of the TEC gradient along each VLA ``arm'' at each antenna.  While directional information is essentially lost within this approach, one may use it to recover information about isotropic fluctuations on scales $<\! 1$ km.  To this end, we have computed a median power spectrum among all antennas from each single-source observation by simply performing a Fourier transform of the projected TEC gradient time series for each antenna and determining the median spectral power at each temporal frequency.  Using the median rather than the mean helps to mitigate the influences of both strong wavelike disturbances and any spurious data that might be present.  These median one-dimensional spectra therefore provide an excellent means to explore the climatological nature of turbulent fluctuations, which is quite complementary to the wave-based analysis provided by the multi-source spectral cubes and single-source, two-dimensional spectral maps.

\section{Climatology}
\subsection{Multi-source 3-D Spectra}
As described above, we used the outputs from the VLSSr field-based calibration procedure for each of the 1862, $\sim\!\! 20$-minute observations to construct fluctuation spectrum cubes, one temporal dimension and two spatial, of the TEC gradient.  We then binned these by local time and then within local time bins, by day of the year, $K_p$ index, and $F10.7$.  We used four time of day bins, dawn (04:00--08:00), daytime (08:00--16:00), dusk (16:00--20:00), and nighttime (20:00-04:00).  We likewise used four day of the year bins, spring (34--124), summer (125--215), autumn (216--306), and winter (307--33).  During the observations, the $K_p$ index varied between 0 and a little more than 5, and we consequently designed our four $K_p$ bins to span the range 0--5.  The values of $F10.7$ ranged from about 70 Solar Flux Units (SFU; 1 SFU $=10^{-22}$ W m$^{-2}$ Hz$^{-1}$) to more than 220 SFU between 2001 and 2007 (i.e., roughly solar maximum to solar minimum).  We therefore constructed four logarithmically spaced bins spanning log $F10.7 = 1.85$--2.35.\par
\begin{figure}
\noindent\includegraphics[width=6in]{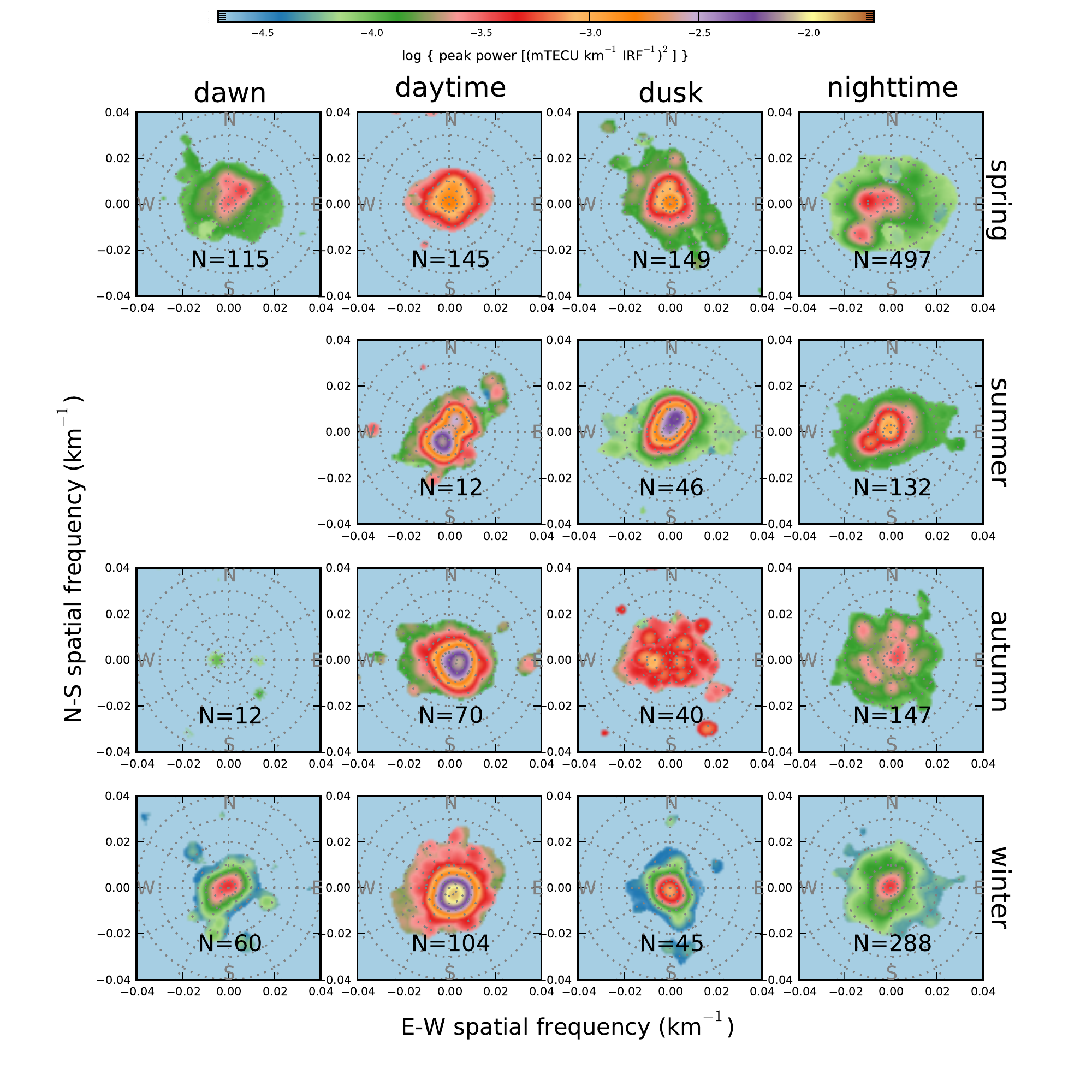}
\caption{From mean fluctuation spectrum cubes, derived from multi-source data and binned by local time and season (see Section 3.1), the peak power over all temporal frequencies as a function of north-south and east-west spatial frequencies.  Locations where the mean power over all temporal frequencies is lower than $5 \times \mbox{MAD}$ are set to zero, where MAD is the median absolute deviation among the mean power values of all spatial frequencies.  The number of observations averaged together to make each mean spectrum is printed in each panel.}
\label{pkspec}
\end{figure}
\begin{figure}
\noindent\includegraphics[width=6in]{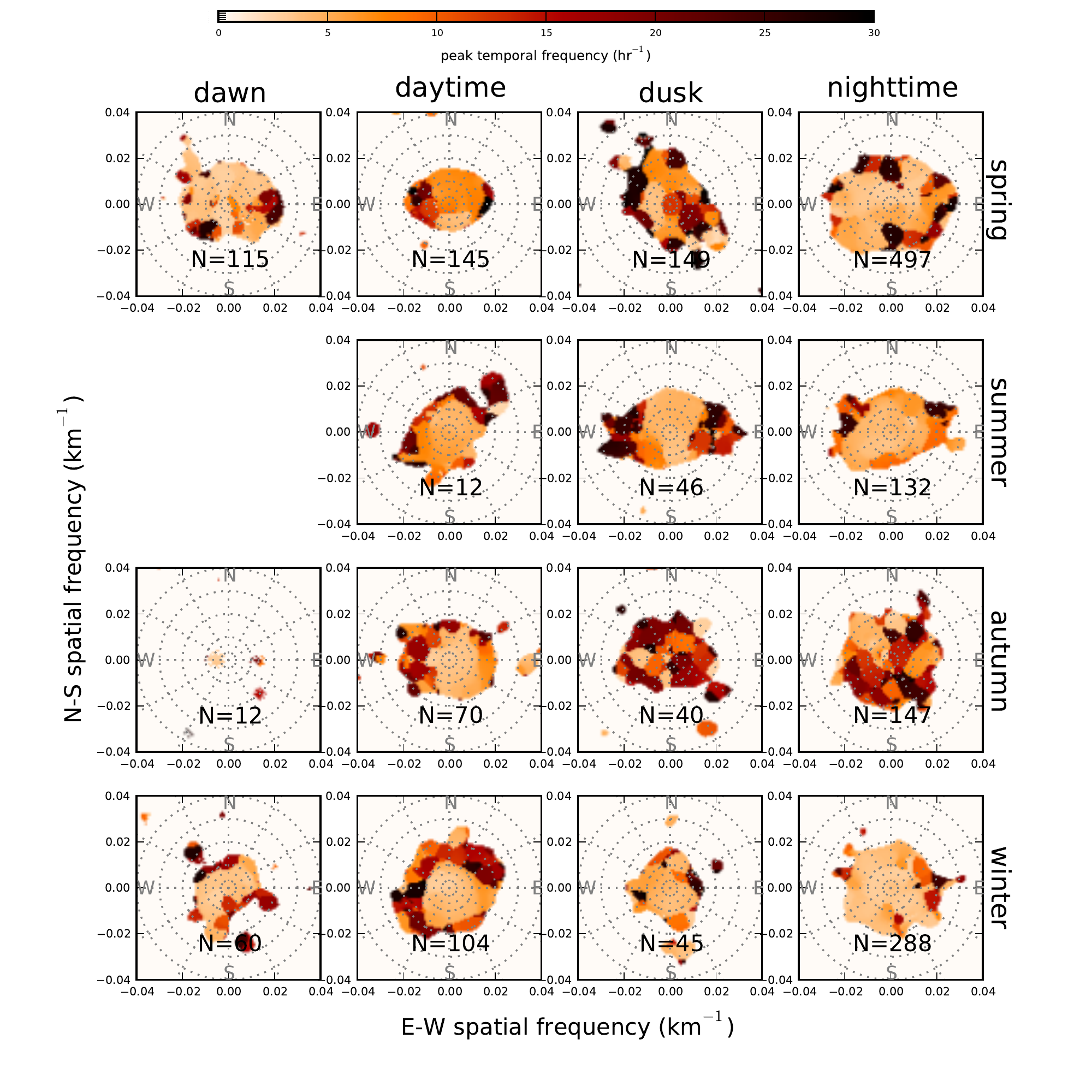}
\caption{From mean fluctuation spectrum cubes, derived from multi-source data and binned by local time and season (see Section 3.1), the peak temporal frequency as a function of north-south and east-west spatial frequencies.  Locations where the mean power over all temporal frequencies is lower than $5 \times \mbox{MAD}$ are set to zero, where MAD is the median absolute deviation among the mean power values of all spatial frequencies. The number of observations averaged together to make each mean spectrum is printed in each panel.}
\label{nuspec}
\end{figure}

In Fig.\ \ref{pkspec}, we display the peak power over all temporal frequencies as a function of $\xi_{NS}$ and $\xi_{EW}$ of the average power spectrum cube for each time-of-day/time-of-year bin.  Maps of the accompanying peak temporal frequency are shown in Fig.\ \ref{nuspec}.  Within each panel, the number of observations used in the average is printed.  For all of the spectra, the full width at half maximum (FWHM) of the IRF is 0.009 km$^{-1}$.  Within the spectral maps, there are a variety of features that are $\gtrsim$ the size of the IRF which likely represent real wavelike structures rather than random fluctuations.  For example, if one examines the spring nighttime panel in Fig.\ \ref{pkspec}, one can see a prominent feature in the lower left quadrant of the plot.  These maps are constructed such that north is up and east is to the right, implying that this feature corresponds to southwestward-directed waves.  There is a polar grid plotted as dotted grey lines to help one estimate direction and wavelength.  In this case, the direction is about $45^\circ$ south of west and the spatial frequency, $\xi$, is about 0.015 km$^{-1}$, implying a wavelength of 67 km (i.e., $1/\xi$).  If one examines the same region in the corresponding panel in Fig.\ \ref{nuspec}, on can see that the peak temporal frequency is about 5 hr$^{-1}$.  Taken with the wavelength, this gives a speed of roughly 330 km hr$^{-1}$, or about 90 m s$^{-1}$.\par
Some of the features apparent within the panels of Fig.\ \ref{pkspec} and \ref{nuspec} seem to be specific to a given season and/or time of day.  Some coincide with the known climatological behavior of previously explored fluctuations.  A summary of prominent features found within these data and the single-source spectral analysis presented below is given in Table \ref{sum}, including comparisons with previous work.\par
As alluded to in Section 1, southwestward-directed MSTIDs have been previously observed to be prominent during summer nighttime in North America.  Likewise, winter daytime MSTIDs are known to be common in the same region, usually propagating toward the southeast \citep[e.g.,][]{her06,tsu07}.  We see direct evidence of these same phenomena within the multi-source spectra.  The summer nighttime spectrum shows a southwest-directed feature with a wavelength of about 100 km and a period of about 17 minutes, implying a speed of 100 m s$^{-1}$.  This is roughly consistent with the known properties of MSTIDs.  Similarly, the winter daytime spectrum shows a southeastward-directed feature with a peak at a wavelength of 190 km, a period of 17 minutes, and an implied speed of 190 m s$^{-1}$, which is again, consistent with MSTIDs.\par
There are also many apparent features within the multi-source spectra that are not commonly known.  For instance, several show evidence of wave activity directed toward the northeast, namely spring dawn, summer daytime and dusk, and autumn dusk and nighttime.  There are southwestward-directed waves during spring nighttime, but they are smaller ($\sim\!\! 50$ km) and slower ($\sim\!\! 60$ m s$^{-1}$) than the MSTIDs seen during summer nighttime.  In addition, the spring nighttime spectrum shows evidence of MSTID-like, westward-directed waves not seen during other time periods.  While medium-scale wave activity seems to be common during the daytime, the predominant direction appears to have a seasonal dependence with southwestward-directed waves being more prevalent during summer daytime and southeastward-directed waves dominating during autumn and winter.  The autumn dusk and nighttime spectra are particularly interesting as they show a plethora of features with no apparent dominant direction or size/speed.  Most of the spectra show evidence of relatively small ($|\xi|\!>\!0.01$ km$^{-1}$) features in the periphery of the areas they occupy within the $\xi_{NS}$,$\xi_{EW}$ plane.\par
\begin{figure}
\noindent\includegraphics[width=6in]{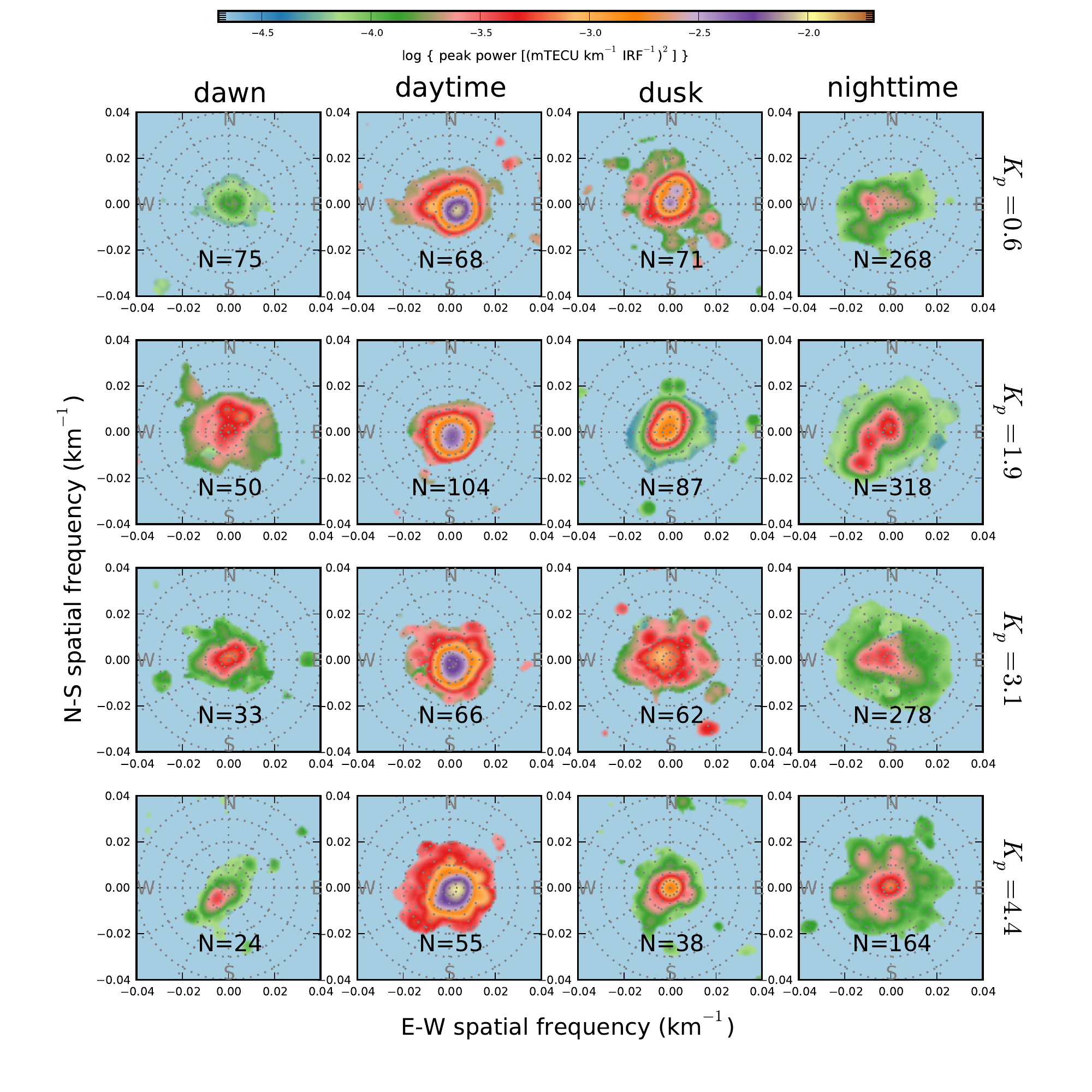}
\caption{The same as Fig.\ \ref{pkspec}, but for spectra binned by local time and $K_p$ index.}
\label{kpspec}
\end{figure}

In Fig.\ \ref{kpspec}, we have plotted spectra similar to those displayed in Fig.\ \ref{pkspec} but for bins of time-of-day and $K_p$ index.  Once can see from these that most of the features seen in Fig.\ \ref{pkspec} do not appear to have a clear dependence on geomagnetic activity.  For instance, southeastward-directed waves are seen during the daytime at all values of $K_p$.  However, it does appear that the northeastward-directed waves that appear at dawn are more prominent at low to moderate values of $K_p$ as are the southwestward-directed nighttime waves (both the $\sim\!\! 50$ and $\sim \!\! 100$ km sized waves).\par
\begin{figure}
\noindent\includegraphics[width=6in]{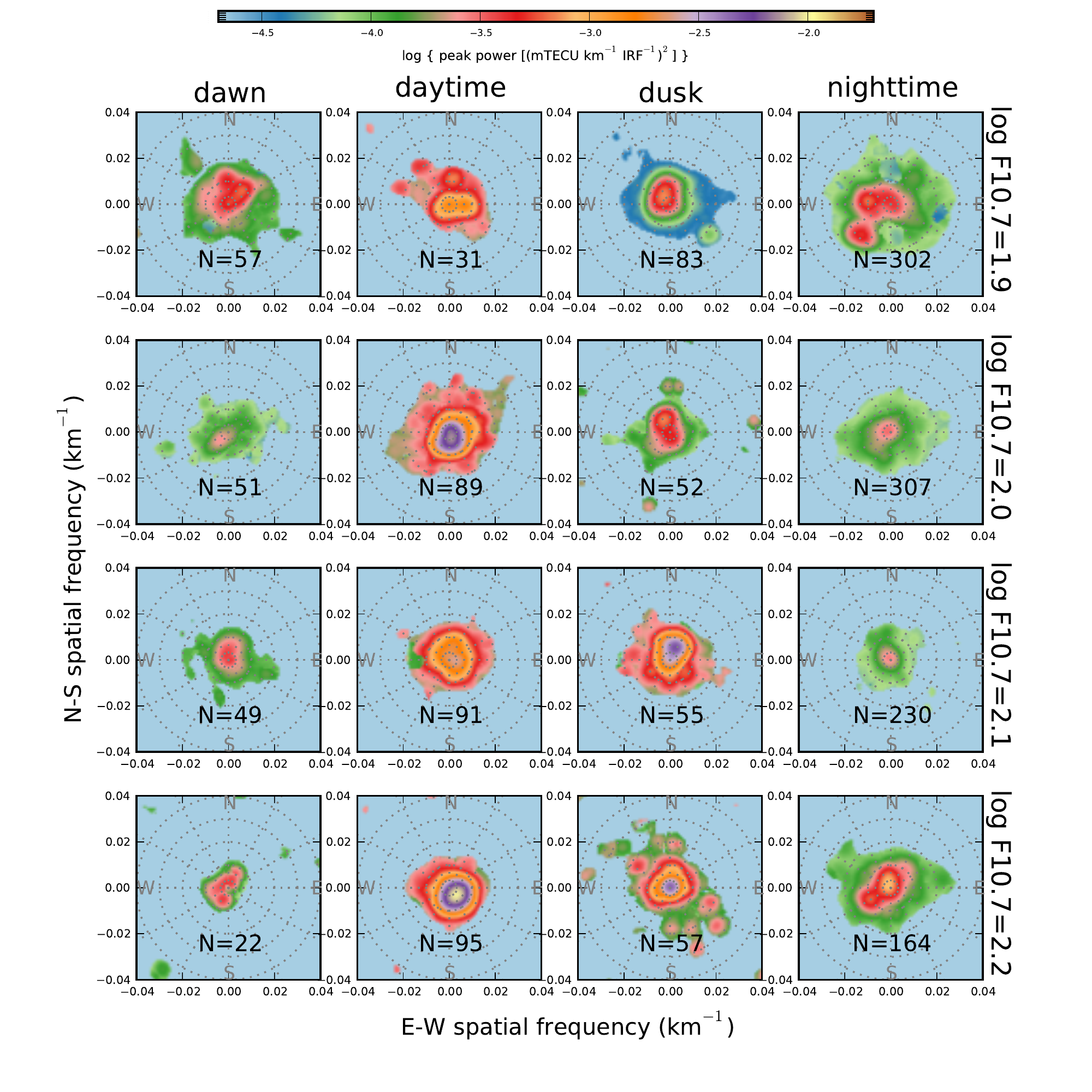}
\caption{The same as Fig.\ \ref{pkspec}, but for spectra binned by local time and $\mbox{log } F10.7$.}
\label{ffspec}
\end{figure}

Similar spectra are plotted again in Fig.\ \ref{ffspec}, this time for bins of time-of-day and $\mbox{log } F10.7$.  Like the results for the $K_p$ index, most of the features seen in Fig.\ \ref{pkspec} do not seem to have noticeable dependences on solar activity.  Again, the most notable exceptions are the dawn northeastward-directed waves and nighttime southwestward-directed waves, both of which are seen most prominently at the lowest levels of solar activity.  The larger southwestward-directed waves seem to actually be most prominent at the highest solar activity levels.  However, we note that all of the summer observations were taken in 2002 near solar maximum.  Indeed, the spectrum in the high $\mbox{log } F10.7$, nighttime bin looks very similar to the summer nighttime spectrum in Fig.\ \ref{pkspec}.  The fact that these waves have been seen by others to have a strong seasonal dependence suggests that it is the time of year rather than solar activity levels that has had the strongest influence on the appearance of this feature in this case.\par
For the lowest $K_p$ and $\mbox{log } F10.7$ bins, the nighttime spectra show a significant eastward-directed feature that is not prominent in the nighttime spectra in Fig.\ \ref{pkspec}.  The feature is actually closer to magnetic east, which is about $12^\circ$ south of due east at the VLA.  The wavelength and speed of this feature are about 150 km and 150 m s$^{-1}$, respectively.  Because these waves are predominantly magnetic eastward-directed with relatively large speeds, they may in fact be field-aligned irregularities within the plasmasphere similar to those discovered with the VLA by \citet{jac92}.  For instance, co-rotation within the plasmasphere will yield an observed speed for such irregularities of 150 m s$^{-1}$ for heights of about 2000 km.  For these observations, this corresponds to a McIlwain $L$-parameter of about 2.1, which is typical for the irregularities previously discovered with the VLA \citep{hoo97}. These disturbances will be discussed further in Section 5.

\subsection{Single-source 2-D Spectra}
\begin{figure}
\noindent\includegraphics[width=6in]{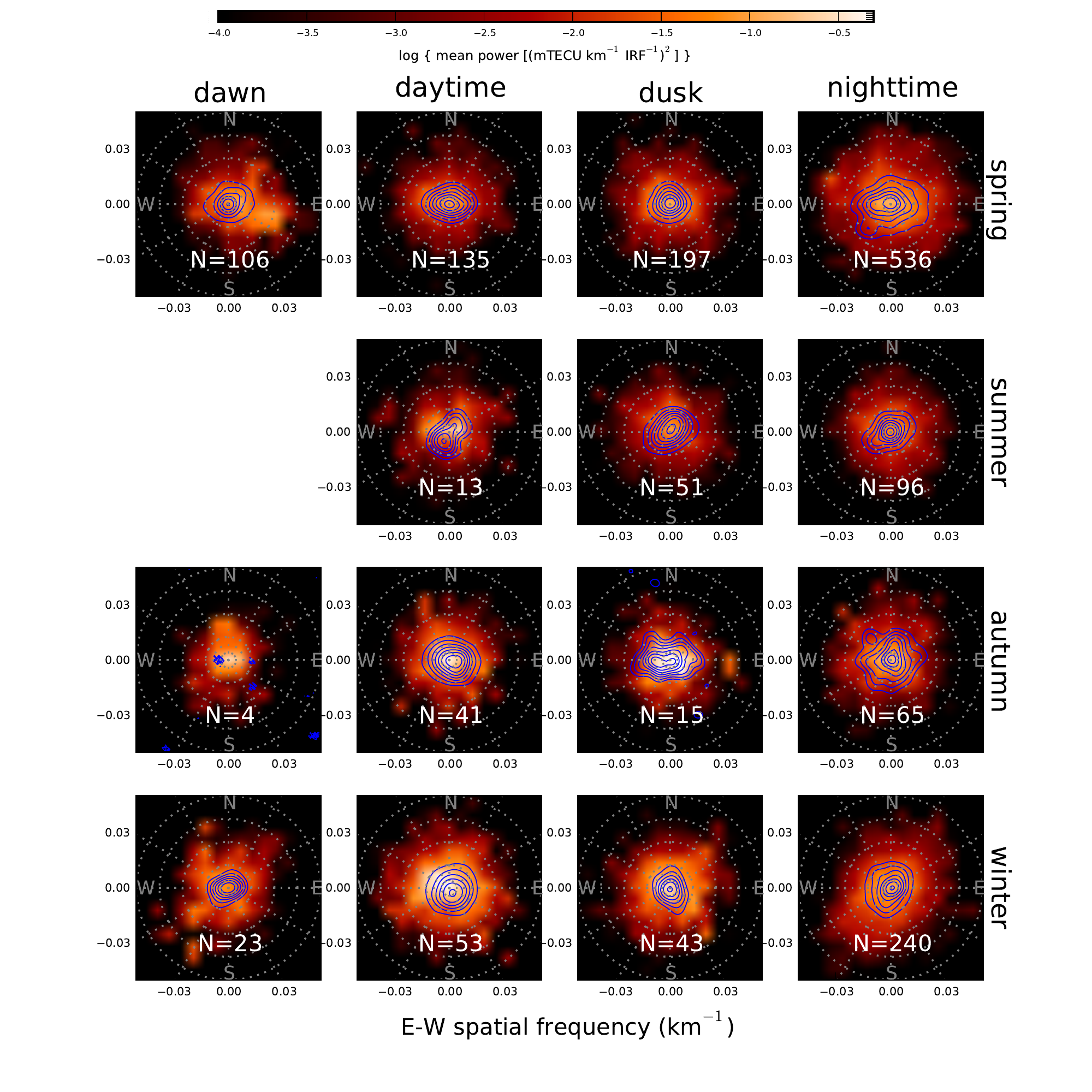}
\caption{Mean maps of TEC gradient fluctuation power, derived from single-source data and binned by local time and season (see Section 3.2).  The blue contours represent the mean power over all temporal frequencies from the corresponding spectral cubes shown in Fig.\ \ref{pkspec} and \ref{nuspec}. The number of observations averaged together to make each mean map is printed in each panel.}
\label{xyimg}
\end{figure}

\begin{figure}
\noindent\includegraphics[width=6in]{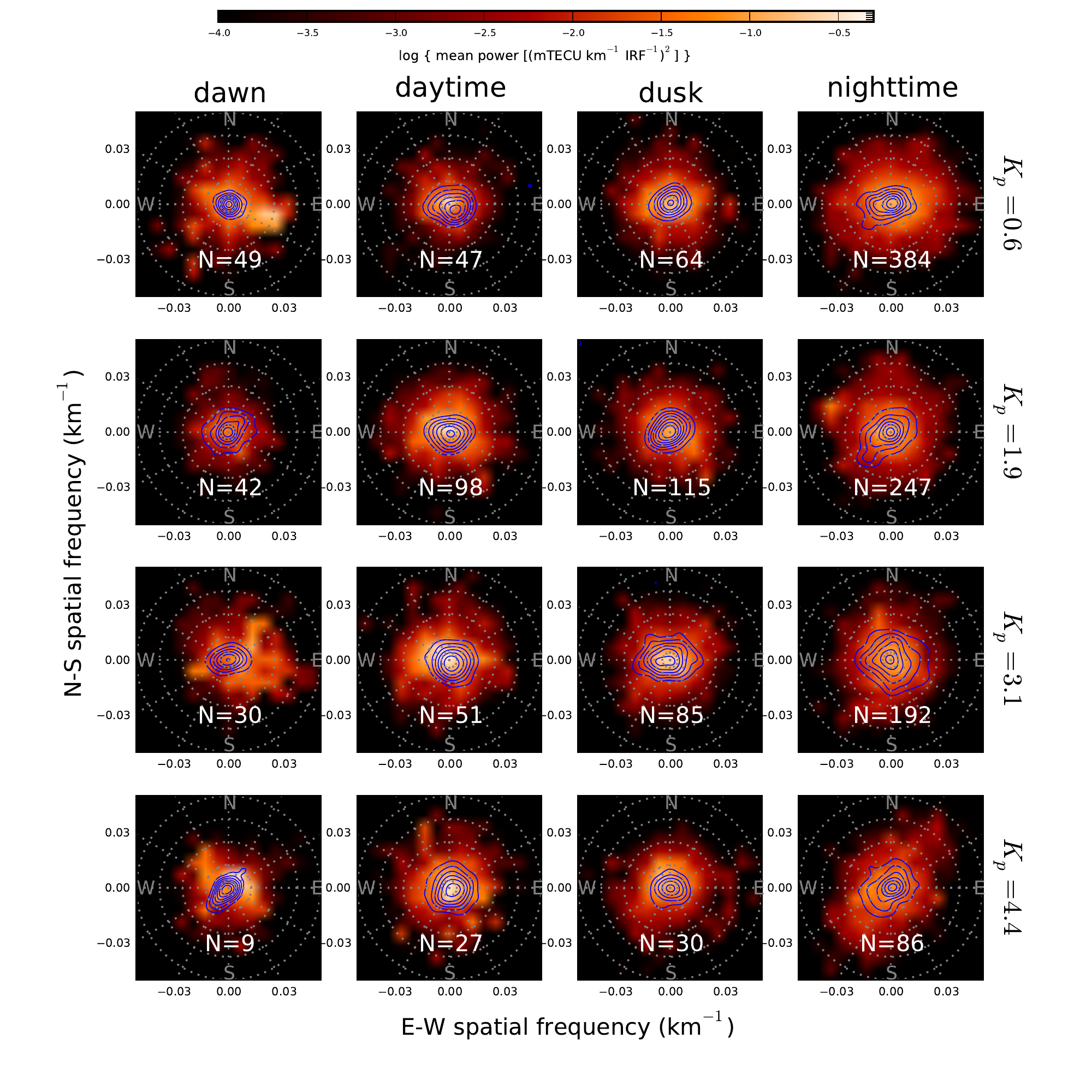}
\caption{The same as Fig.\ \ref{xyimg}, but for spectra binned by local time and $K_p$ index.}
\label{kpimg}
\end{figure}

\begin{figure}
\noindent\includegraphics[width=6in]{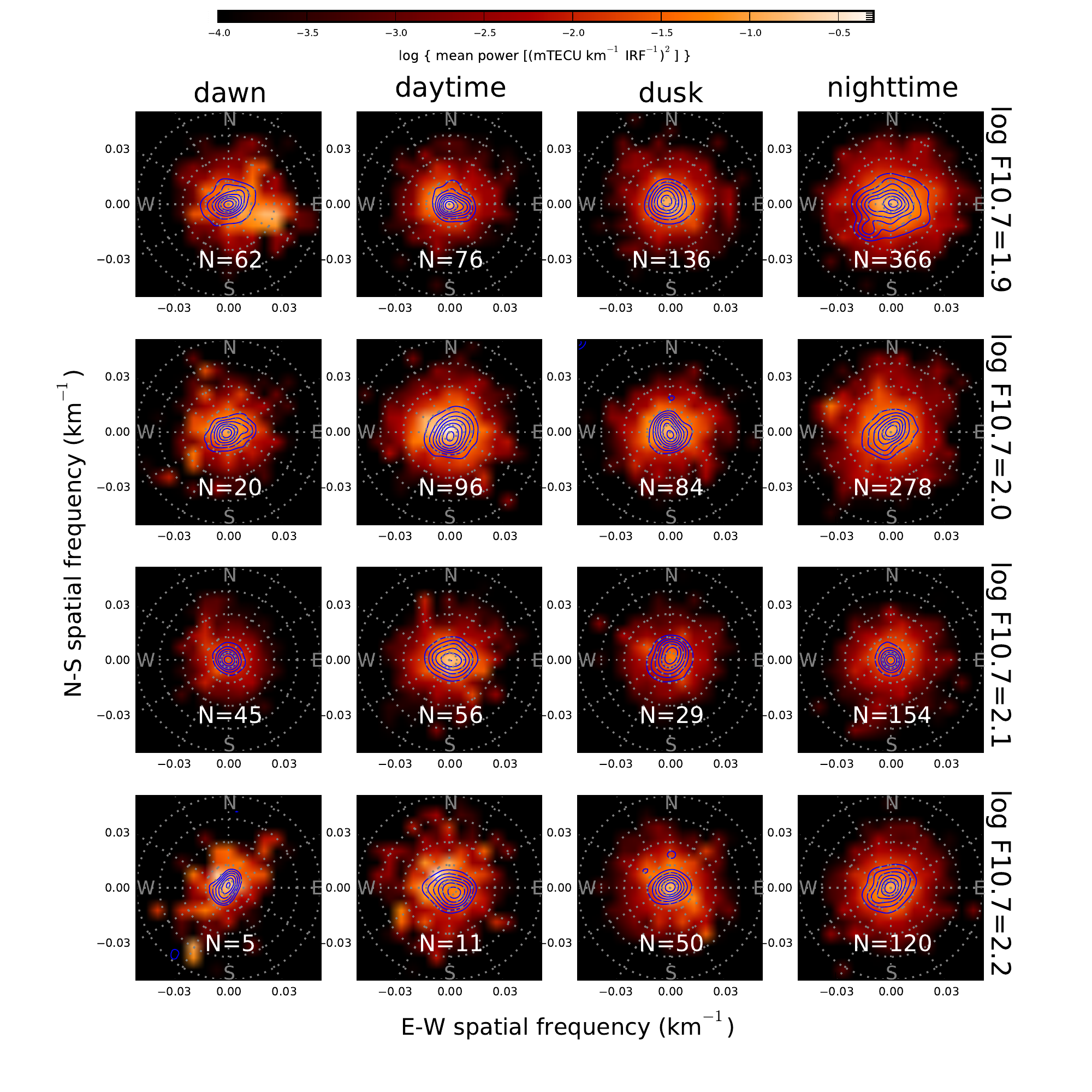}
\caption{The same as Fig.\ \ref{xyimg}, but for spectra binned by local time and $\mbox{log } F10.7$.}
\label{ffimg}
\end{figure}

As discussed in Section 3.2, the spectral analysis of each single-source observation yielded a measurement of the distribution of spectral power in the $\xi_{NS}$,$\xi_{EW}$ plane that reaches higher spatial frequencies than the multi-source spectra.  We have binned these spectral maps according to local time, season, $K_p$ index and $F10.7$ in the same way as the multi-source spectra and have displayed the results in Fig.\ \ref{xyimg}--\ref{ffimg}.  In each panel of each of these figures, we have plotted the mean power over all temporal frequencies from the corresponding mean multi-source spectral cube as blue contours.\par
The spectral maps shown in Fig.\ \ref{xyimg} show that at all times of day and year, there are significant fluctuations at scales too small to detect with the multi-source data.  However, when averaged together, they seem to be largely isotropic with a few notable exceptions.  First, the map for winter daytime seems to be especially asymmetric, showing an over-density in the northwest quadrant.  This northwest over-density is dominant on scales of about 100 km.  Fig.\ \ref{kpimg} and \ref{ffimg} show that the northwest over-density seen during winter daytime is more prominent at moderate solar activity levels, but does not seem to have a dependence on $K_p$.\par
The autumn dawn map shows a population of nearly northward-propagating disturbances with wavelengths of about 50 km not seen in the corresponding multi-source spectrum.  There is some indication from Fig.\ \ref{ffimg} that these waves occur more frequently during times of relatively high solar activity.\par
Perhaps the most striking feature is a significant group of small-scale waves directed toward the southeast during spring dawn with wavelengths of about 40 km which represent a distinct population of waves not detectable with the multi-source data.  Fig.\ \ref{kpimg} and \ref{ffimg}  show that these waves also appear prominently at times of low geomagnetic and solar activity.\par
\begin{figure}
\noindent\includegraphics[width=6in]{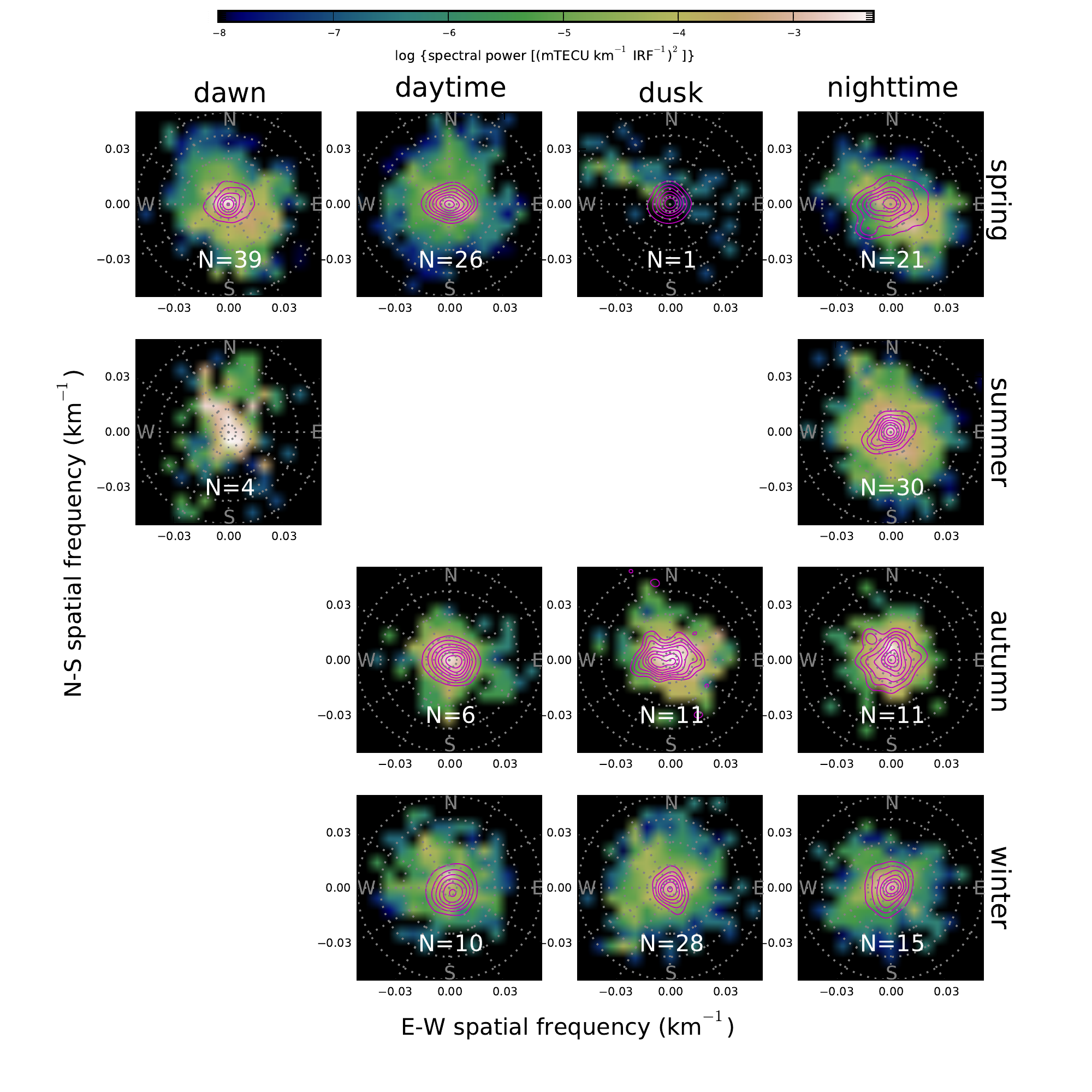}
\caption{Mean maps of TEC gradient fluctuation power, derived from five-minute observations of the bright calibration source, Cygnus A.  These maps were made by binning the observations by local time and season (see Section 3.2).  The magenta contours represent the mean power over all temporal frequencies from the corresponding spectral cubes shown in Fig.\ \ref{pkspec} and \ref{nuspec}. The number of observations averaged together to make each mean map is printed in each panel.}
\label{xycyg}
\end{figure}
\begin{figure}
\noindent\includegraphics[width=6in]{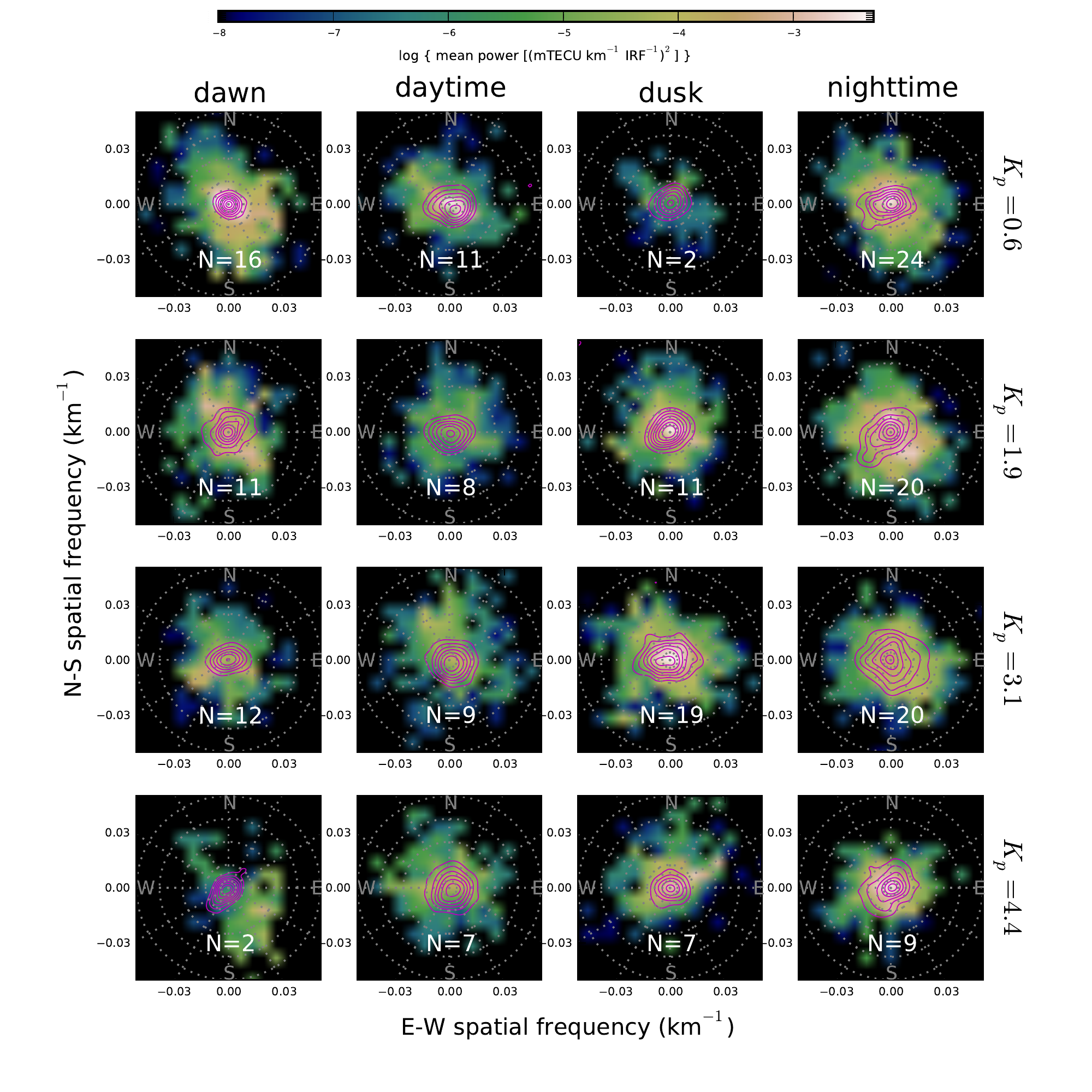}
\caption{The same as Fig.\ \ref{xycyg}, but for spectra binned by local time and $K_p$ index.}
\label{kpcyg}
\end{figure}
\begin{figure}
\noindent\includegraphics[width=6in]{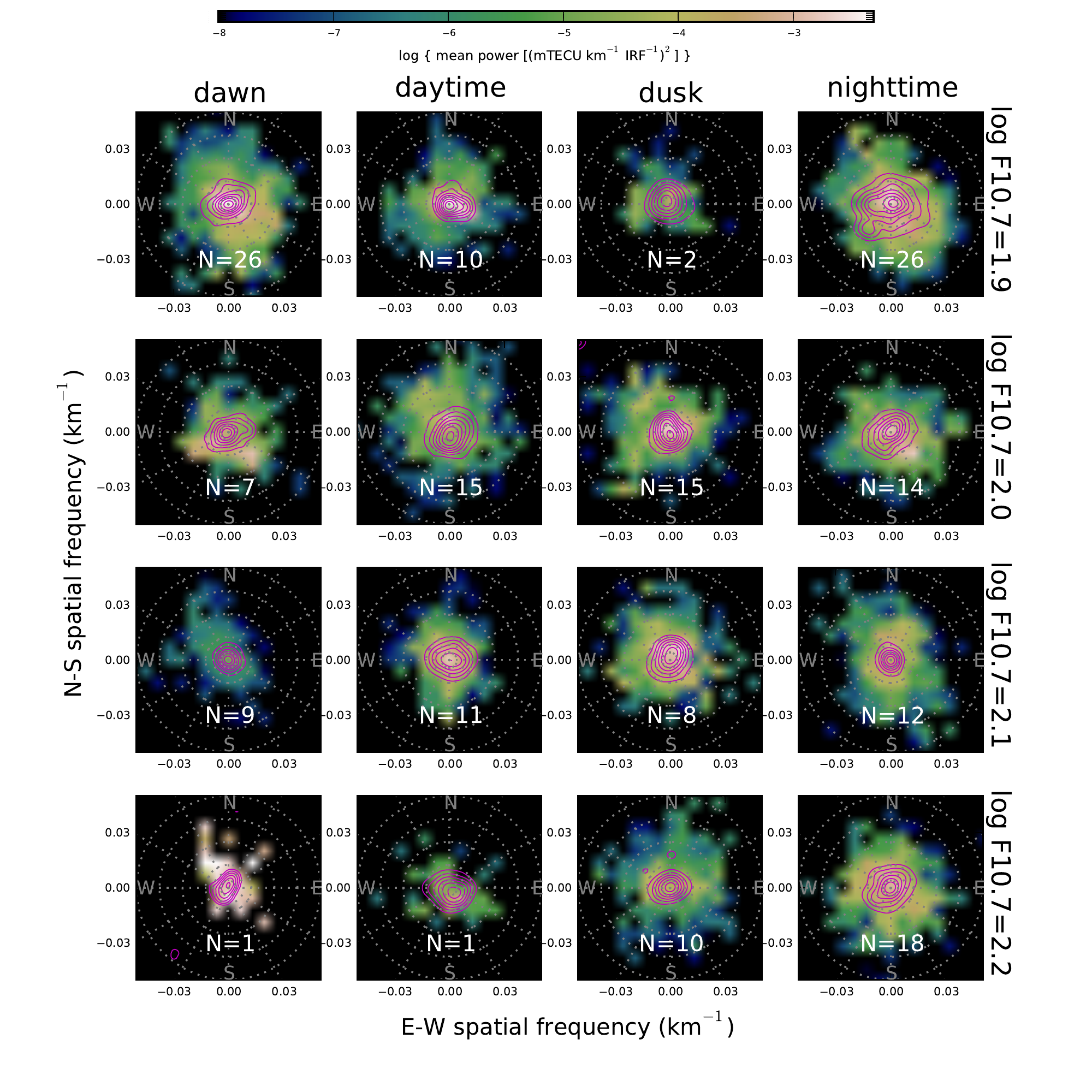}
\caption{The same as Fig.\ \ref{xycyg}, but for spectra binned by local time and $\mbox{log } F10.7$.}
\label{ffcyg}
\end{figure}

We also discussed in Sections 2.3 and 3.2 separate five-minute calibration observations that were made of the extremely bright source, Cyg A, which we also used to produce spectral maps.  There were significantly fewer of these observations (202 versus 1618 in the B configuration), and their time-of-day and seasonal coverage is not as good.  However, as also discussed in Section 3.2, the spectral maps produced from these observations are more sensitive to smaller-scale fluctuations, making them complementary to the single-source maps shown in Fig.\ \ref{xyimg}--\ref{ffimg}.  Therefore, we have displayed in Fig.\ \ref{xycyg}--\ref{ffcyg} Cyg A-based spectral maps binned by local time and season (Fig.\ \ref{xycyg}), $K_p$ index (Fig.\ \ref{kpcyg}), and $F10.7$ (Fig.\ \ref{ffcyg}).\par
The Cyg A-based spectra show no signs of any significant groups of waves.  However, as expected, they typically extend to larger spatial frequencies than their counterparts shown in Fig.\ \ref{xyimg}--\ref{ffimg}.  While the maps often show roughly isotropic distributions of fluctuations, reminiscent of turbulence, many show some evidence of ``preferred'' directions.  For instance, the spring nighttime map is elongated along the northwest-southeast axis with more power apparent in the southeast direction.  The same appears to be true during dawn for higher $K_p$ indices, but with a much larger fraction of the total spectral power in the southeastern quadrant.  The summer dawn spectrum shows a significant northwestern extension toward higher spatial frequencies.\par
The summer dawn map has an extension toward the northwest that reaches spatial frequencies as high as about 0.04 km$^{-1}$ (or, scales of 25 km).  Similar features can be seen in the dawn maps in Fig.\ \ref{ffcyg} for moderate to high levels of solar activity.  There is also a prominent group of northwestward-directed waves extending to large (0.05 km$^{-1}$) spatial frequencies in the spring dusk map, but this is based on a single observation.

\subsection{Single-source 1-D Spectra and Turbulence}
As described in Section 3.2, we have produced median TEC gradient fluctuation spectra using the arm-based method of computing the projected TEC gradient from all single-source data.  Since the projected gradients do not rely on polynomial fits, we have used all 1978 single-source observations for this analysis, including those conducted in the hybrid BnA configuration.  These one-dimensional spectra probe the shape of the spectrum of isotropic of fluctuations down to the smallest periods possible, 40 seconds.  If we assume that the population of these fluctuations does not change significantly during each observation, the temporal frequencies can be directly related to spatial ones using the apparent speed of the observed cosmic source at ionospheric heights.  Since turbulence is likely to be much more prominent at lower altitudes where ion-neutral coupling is more important, we have computed the median apparent source speed at the height of the $E$-region, 100 km, which is roughly 10 m s$^{-1}$.  This implies that these spectra probe scales as small as 0.4 km.\par
\begin{figure}
\noindent\includegraphics[width=6in]{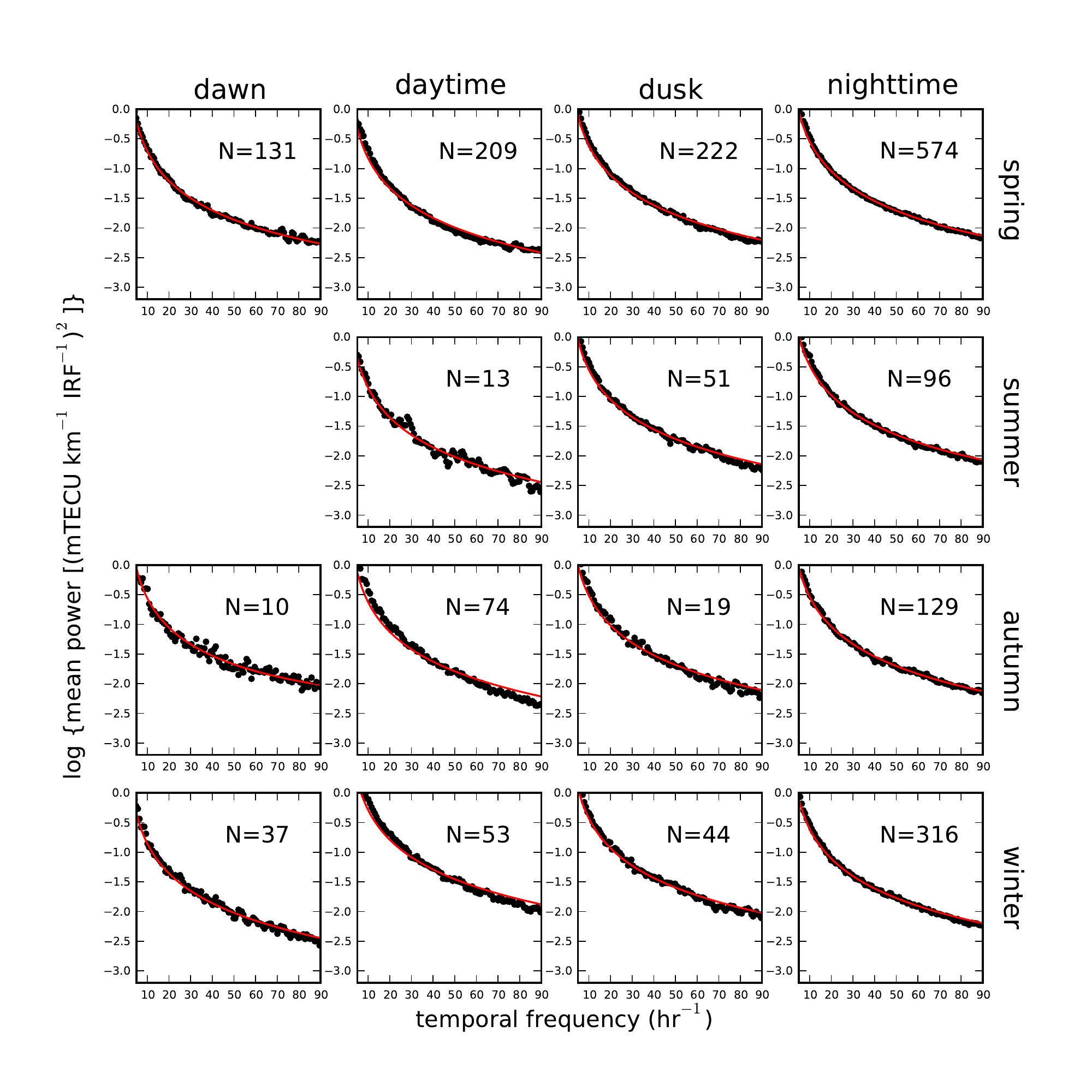}
\caption{Within bins of local time and season, the mean one-dimensional fluctuation spectra derived from single-source data using the arm-based method described in Section 2.2.  The turbulence model described in Section 3.3 was fit to each of these spectra and is plotted as a red curve.}
\label{arm}
\end{figure}

Continuing with the assumption of ``frozen'' turbulence, we can characterize the level of turbulent activity by fitting a simple power-law model to each spectrum.  Specifically, following the work of \citet{kol41a,kol41b} and \citet{tat61}, the turbulent spectrum of phase fluctuations, which in this case is the same as TEC fluctuations, should be proportional to $\xi^{-11/3}$.  Therefore, the TEC gradient spectrum should be $\propto \xi^{-5/3}$ and the temporal spectra should be well fit by a spectrum $\propto \nu^{-5/3}$ for frozen turbulence.  We have displayed mean one-dimensional single-source spectra in Fig.\ \ref{arm}, binned by season and local time, to illustrate that this is generally the case.  For this analysis, we have excluded the five-minute Cyg-A observations because the shorter duration yields lower temporal spectral resolution which significantly alters the shapes of the spectra at lower frequencies.  We also found that the 20-minute, single-source observations were more than adequate for characterizing the spectrum of turbulent fluctuations (see Fig.\ \ref{arm}) and that the Cyg A-based spectra did not significantly improve or add to this analysis.\par
For each single-source spectrum, we have fit a simple model of $P_T \nu^{-5/3} + N$, where $P_T$ is a parameter that characterizes the level of turbulent activity and $N$ is a constant that models the effect of noise within the spectrum.  The fits were constrained to $\nu\!\! > 20$ hr$^{-1}$ because below this limit, the influence of the window function used (i.e., a simple 20-minute wide boxcar) as well as relatively strong, medium-to-large scale fluctuations on the shape of the spectrum can be significant.  The fits are plotted as red curves in Fig.\ \ref{arm}.  One can see that these turbulence model fits provide a good approximation of the data in all cases.\par
\begin{figure}
\noindent\includegraphics[width=6in]{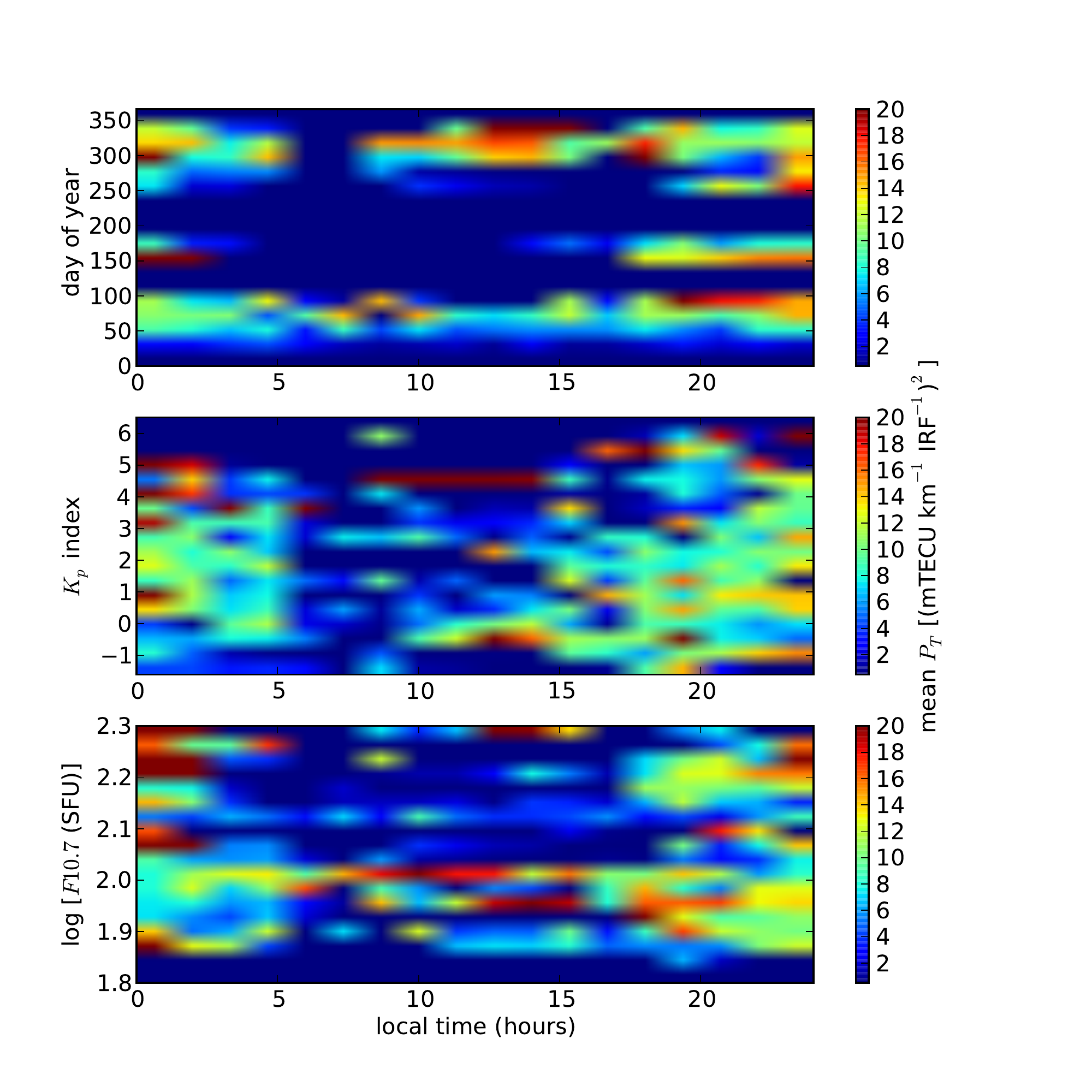}
\caption{The strength of the turbulent fluctuation power, $P_T$ (see Section 3.3), derived from single-source data as a function of local time and day of the year (upper), $K_p$ index (middle), and $F10.7$ (lower).}
\label{turb}
\end{figure}

To obtain a climatological picture of turbulent activity, we also fit the turbulence model described above to the arm-based spectrum from each single-source observation.  Among these spectra, the parameter $P_T$ varied between roughly 1 and 20 (mTECU km$^{-1}$ IRF$^{-1}$)$^2$ and the median value for the parameter $N$ was $10^{-4}$ (mTECU km$^{-1}$ IRF$^{-1}$)$^2$.  The mean value of $P_T$ was then computed within bins of local time and day of the year, $K_p$ index, and $\mbox{log } F10.7$.  These two-dimensional distributions are displayed in the panels of Fig.\ \ref{turb}.  During the day, the turbulent activity appears to be highest during winter months, which is when daytime MSTID activity also peaks.  During dusk, it seems to be higher in the late winter/early spring while post-midnight and before dawn, there seems to be some increased activity during the summer.  There also seems to be a ``spike'' of turbulent activity during the day at moderately high values of $K_p$ ($4+$ to $5-$).  Daytime turbulent activity seems to be relatively high during moderate levels of solar activity ($F10.7\sim \!\! 100$ SFU).  Post-midnight and before dawn, the turbulent activity also has peaks near $F10.7\! \approx \!120$ and 170 SFU.

\section{Discussion}
We have used a large, 74 MHz survey of the northern sky, the VLSS, to explore the rich environment of ionospheric disturbances over the southwestern United States and northern Mexico (see Fig.\ \ref{cover}).  Some of the results were expected from previous analysis, namely the prominent MSTIDs during summer nighttime and winter daytime, which propagate toward the southwest and southeast, respectively \citep[see, e.g.,][]{her06,tsu07}.  However, there are several features unique to this study, and we discuss a subset of them below.
\subsection{Muti-directional Waves}
The results shown in Fig.\ \ref{pkspec} show evidence of waves moving in several different directions on different scales.  Most notable are the results for autumn dusk and nighttime, each showing this variety within a single mean spectrum.  While several factors may influence the appearance of these waves, it seems most likely that they are related to orographic gravity waves \citep[see, e.g.,][]{vad10}.  This is because beneath the region of the ionosphere probed by the VLSS observations, the terrain is quite mountainous.  The is illustrated within the bottom panel of Fig.\ \ref{cover} where one can see from the displayed relief map that the latitudes and longitudes of the ionospheric pierce-points for the VLSS observations cover the most mountainous regions of northern Mexico, New Mexico, and Arizona.  In addition, the Rocky Mountains lie largely to the north and west of these observations, implying that features like the southeastward-directed, 40-km sized waves seen during spring dawn may be related to waves generated by airflow over the Rockies.
\subsection{Northeastward Waves}
A prominent and repeating feature within the spectra shown in Fig.\ \ref{pkspec} is a class of northeastward-directed waves.  These are similar in size, and often in strength, to MSTIDs, but are traveling in a relatively abnormal direction for these types of disturbances.  Similar waves were observed with the VLA during nighttime observations in August, 2003 by \citet{hel12b}.  The relief map in Fig.\ \ref{cover} does indicate that there are mountains to the southwest of many of the ionospheric pierce-points, implying that gravity waves may be involved.  However, these northeastward-directed waves are often relatively strong compared to other spectral features seen in Fig.\ \ref{pkspec} (see spring dawn and summer daytime for examples).  The mountains to the southwest are not particularly large, especially when compared to the Rockies to the northwest, implying that there may be more at work here then vertically propagating gravity waves.  The northeastward-directed waves detected by \citet{hel12b} were seen to coincident with drops in $F$-region height measured with relatively nearby ionosondes data.  This is similar to what was observed over Japan by \citet{shi08}.  It is therefore plausible that $F$-region compressions are related to the appearance of these features within the VLSS data.  However, a thorough investigation using contemporaneous ionosondes data is required to resolve this and is currently underway.
\subsection{Magnetic Eastward Waves}
As noted in Section 3.1, there are relatively unique, eastward-directed features in the nighttime spectra for the lowest $K_p$ and $\mbox{log } F10.7$ bins shown in Fig.\ \ref{kpspec} and \ref{ffspec}.  These could be similar to the magnetic eastward-directed waves discovered with the VLA and established by \citet{hoo97} to be plasmaspheric waves.  They could also be associated with disturbances within plasma flow from the plasmasphere to the ionosphere discovered with the VLA by \citet{helint}.  The latter seems particularly likely because (1) they are only seen during the night when such flows are most likely to occur and (2) the spectra also show evidence of westward-directed, presumably ionospheric waves which were also found by \citet{helint} to occur simultaneously with the plasmaspheric disturbances.  The fact that they are most prominent during periods of low geomagnetic and solar activity may imply that a relatively undisturbed plasmasphere is required to produce these features.  A more detailed study of this phenomenon is currently being conducted using GPS-based TEC maps, $K_p$ and AE indices, and VLSS data.  The preliminary results suggest that these disturbed flows may be triggered by localized depletions within the nighttime ionosphere triggered by forcing from the lower atmosphere, likely related to tides.  The full results will be presented in a subsequent paper.
\subsection{Turbulence}
The results shown in Fig.\ \ref{arm} confirm what was demonstrated by \citet{coh09} and \citep{hel12b}, that the median behavior of the spectrum of ionospheric fluctuations is turbulent.  Fig.\ \ref{turb} shows that while the level of turbulence does vary over time, the difference between the maximum and minimum levels is only about a factor of five.  Contrast this with the results for wave activity displayed in Fig.\ \ref{pkspec}--\ref{ffcyg} where the range in spectral power spans at least two orders of magnitude.  The level of turbulent activity does appear to be related in some way to the occurrence of certain wave phenomena.  In particular, $P_T$ is relatively large when MSTIDs are prominent in winter daytime and summer nighttime.\par

\section{Conclusions}
By using a large database of 74 MHz observations of cosmic sources, we have been able to probe the climatological behavior of a variety of ionospheric disturbances over the southwestern United States and parts of Mexico.  We have shown that small-scale, turbulent fluctuations are present nearly all of the time with only a weak dependence on time of day, time of year, and/or the level of geomagnetic and solar activity.  We found a variety of small (roughly 40 km wavelength) and medium (100 km and larger) scale phenomena present at different times.  The sources of these disturbances may range from gravity waves, plasmaspheric interactions, coupling between different ionospheric layers, and plasma instabilities (e.g., the Perkins instability).  Several of these wavelike irregularities have relatively small amplitudes too weak to detect with other methods such as GPS-based TEC measurements.  In some cases, the wave activity was coincident with increases in turbulent activity, namely during winter daytime and summer nighttime.\par
\begin{figure}
\noindent\includegraphics[width=6in]{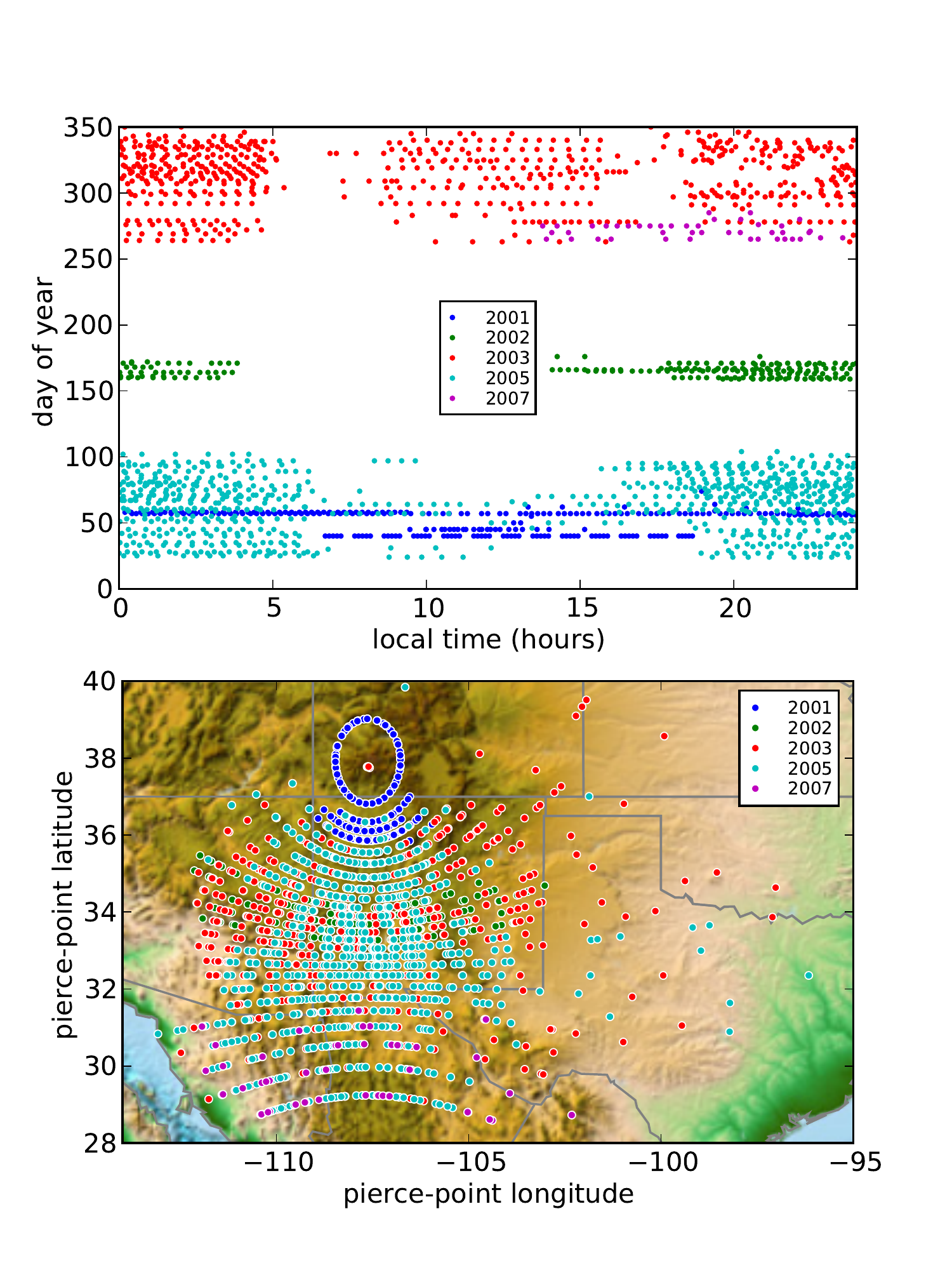}
\caption{Upper:  The local time and day of the year of each of the VLSS observations; points are color-coded by year.  Lower:  The latitude and longitude of the ionospheric pierce-point for the center of the field of view for each VLSS observation assuming a height of 300 km.  The same color-coding by year is used as in the upper panel.  A relief map from $\mbox{http://www.ngdc.noaa.gov/mgg/global/global.html}$ is also displayed for reference.}
\label{cover}
\end{figure}

We note that in general, the impact of the results presented is somewhat hampered by the limited seasonal and time-of-day coverage of the VLSS shown in the upper panel of Fig.\ \ref{cover}.  For instance, all of the summer observations were conducted in the month of June.  The fact that the summer nighttime MSTIDs appear weaker than their winter daytime counterparts may be heavily influenced by the lack of data for July and August.  In addition, the only observations conducted during the dawn hours in the summer were calibration observations of Cyg A.  We are now designing a survey with the new 330 MHz system for the VLA, which is currently being commissioned.  This new survey will be scheduled to optimize both time-of-day and seasonal coverage.  It will benefit from two new arrays of GPS receivers operating continuously within New Mexico which have recently increased the number of such stations from $\sim\!\! 6$ during the time of the VLSS to nearly 40.  There will also be a new digital ionosonde operating in nearby Kirtland Air Force Base in Albuquerque which will potentially provide information about the height where detected disturbances occur.  Thus, we will have a contemporaneous GPS-based account of the largest fluctuations, providing a complete and unique inventory of the seasonal dependence of ionospheric disturbances that will expand upon the work presented here.

\par
\begin{acknowledgments}
Basic research in astronomy at the Naval Research Laboratory is supported by 6.1 base funding.  The Very Large Array is operated by The National Radio Astronomy Observatory.  The National Radio Astronomy Observatory is a facility of the National Science Foundation operated under cooperative agreement by Associated Universities, Inc.
\end{acknowledgments}

\end{article}

\clearpage
\begin{table}
\caption{Summary of Observed Phenomena}
\begin{tabular}{lccc}
\hline
Phenomenon & Season(s) & Time(s) of Day & Notes \\
\hline
SW-directed MSTIDs & summer & daytime--nighttime & previously detected with GPS and airglow imagers; e.g., \\
 & & & \citet{her06,tsu07} \\
 & & & \\
SE-directed MSTIDs & autumn/winter & daytime & same as above \\
 & & & \\
NE-directed MSTIDs & spring/summer & dawn/daytime--dusk & isolated cases detected previously by \\
 & & & \citet{shi08} and \citet{hel12b} \\
 & & & \\
 multi-directional, & autumn & dusk--nighttime & likely associated with gravity waves \\
  medium-scale waves & & & \\
  & & & \\
SW- and W-directed & spring & nighttime & seen with magnetic east-directed waves at low $K_p$; \\
medium-scale waves & & & isolated case detected by \citet{helint} \\
 & & & \\
SE-directed 40-km- & spring & dawn & found when $K_p$ is low; \\
scale waves & & & direction close to magnetic east \\
 & & & \\
 turbulence & spring/summer/ & dusk/nighttime/ & expands on diurnal variation \\
 & winter & daytime & seen by \citet{coh09} \\
 \hline
\end{tabular}
\label{sum}
 \end{table}

\clearpage

\clearpage

\clearpage

\clearpage

\clearpage

\clearpage

\clearpage

\clearpage

\clearpage

\clearpage

\clearpage

\clearpage

\clearpage

\clearpage

\end{document}